\documentclass{aa}  
\usepackage{graphicx}
\usepackage{txfonts}
\begin{document}

\title{ SMEI observations of previously unseen pulsation frequencies in $\gamma$ Doradus }

\author{
	N. J. Tarrant\inst{1} \and 
	W. J. Chaplin\inst{1} \and
	Y. P. Elsworth\inst{1} \and
	S. A. Spreckley\inst{1} \and
	I. R. Stevens\inst{1}
}

\offprints{N. J. Tarrant \email{njt@bison.ph.bham.ac.uk} }

\date{}

\institute{
	$^{1}$ School of Physics and Astronomy, University of Birmingham,
    Edgbaston, Birmingham, B15 2TT
}
 
%\address{$^1$ School of Physics and Astronomy, University of Birmingham, Edgbaston, B15 2TT, United Kingdom }

\abstract
% Context
	{}
% Aims
	{As g-mode pulsators, gamma-Doradus-class stars may na\"ively be expected to show a large number of modes. Taking advantage of the long photometric time-series generated by the Solar Mass Ejection Imager (SMEI) instrument, we have studied the star gamma Doradus to determine whether any other modes than the three already known are present at observable amplitude.}
% Methods
	{High-precision photometric data from SMEI taken between April 2003 and March 2006 were subjected to periodogram analysis with the PERIOD04 package.}
% Results
	{We confidently determine three additional frequencies at 1.39, 1.87, and 2.743 d$^{-1}$. These are above and beyond the known frequencies of 1.320, 1.364, and 1.47 d$^{-1}$}
% Conclusions
	{Two of the new frequencies, at 1.39 and 1.87 d$^{-1}$, are speculated to be additional modes of oscillation, with the third frequency at 2.743$^{-1}$ a possible combination frequency.}


   \maketitle
%
%________________________________________________________________

\section{Introduction}

Gamma-Doradus-class variable stars are usually F-type main-sequence or sub-giant stars, located between the Sun-like and delta Scuti oscillators, with some stars of the class showing both gamma-Doradus and delta-Scuti-type oscillations (e.g. Henry and Fekel, 2005). A recent review of current research on the gamma Doradus phenomenon, both theoretical and observational, can be found in Kaye (2007). Gamma-Doradus-class oscillations are  gravitationally restored (g mode) pulsations with frequencies in the range of 0.5 to 3 d$^{-1}$ and amplitudes of a few percent.

As gamma-Doradus-class stars are g-mode pulsators, a great number of modes of oscillation is possible. However, due to the periods lasting about one day, obtaining good data from the ground is challenging, and currently only a few frequencies have been observed in any single star (Uytterhoeven et al., 2008). 
In addition, the pulsation periods are comparable in size to the expected rotational periods, making it difficult to detect and identify the individual pulsation modes (Handler, 2005). Long-period datasets, such as the 150-day observations of CoRoT (Baglin et al., 2006), and potentially longer time-series of Kepler (Christensen-Dalsgaard et al., 2007) promise to enhance our knowledge of these stars. Here we make use of long time-series observations by another satellite instrument, SMEI (Eyles et al., 2003), to investigate the oscillations of gamma Doradus.

Gamma-Doradus-class stars are of particular interest for asteroseismology, as g modes are sensitive probes of stellar cores. The pulsational mechanism for gamma-Doradus-class oscillations is thought to be a flux modulation induced by the upper convective layer (for instance, see Dupret et al, 2006), however, the exact details of this process are still in need of clarification. 

There are currently three known modes of oscillation in gamma Doradus, at frequencies of 1.32098(2), 1.36354(2) and 1.475(1) d$^{-1}$ (Balona et al., 1994). Through photometric methods Dupret et al. (2006) identified the first two of these as $l=1$ modes. The observation and identification of further modes should allow for a more accurate description of the stellar interior. 

The SMEI instrument, aboard the Coriolis satellite, consists of three cameras, each with a sixty-by-three degree field of view. Taken together these give an instantaneous field of view of a 170-by-3 degree strip of the zenith facing hemisphere. This slice of the sky will rotate as the satellite orbits, providing a near complete view of the sky over the course of an entire orbit. A single observation of each star is made once per 100 minute orbit, resulting in a time series with a 100 minute cadence.

Photometric data from SMEI have already been used for asteroseismology; in the detection of Sun-like oscillations in Red Giant stars (Tarrant et al., 2007, 2008) and in studying the period and amplitude evolution of Polaris (Spreckley et al., 2008). The reader is referred to Spreckley and Stevens (in prep.) for a detailed description of the data extraction and processing pipeline. 

\section{Data}

\begin{table}
\centering
\caption{Observation windows of gamma Doradus. Times are given as the Julian date less $2\,450\,000$ (JD-2450000).}
\begin{tabular}{lccccc}
\hline\hline
\multicolumn{6}{c}{Camera \#1} \\
	&	Start	&	End	&	Duration (d)	&	No. Points	&	Fill (\%)	\\
\hline
1	& 2958.71	&	2986.09	& 27.38 & 182 & 47 \\
2	& 3319.60	&	3349.10	& 29.50 & 195 & 47 \\
3	& 3683.93	&	3717.73	& 33.80 & 280 & 58 \\
\hline
\multicolumn{6}{c}{Camera \#2} \\
	&	Start	&	End	&	Duration (d)	&	No. Points	&	Fill (\%)	\\
\hline
1	& 2847.15	&	3092.07	& 244.92 & 1798	& 52 \\
2	& 3215.03	&	3457.68	& 242.65 & 2305 & 67 \\
3	& 3578.45	&	3816.36	& 237.91 & 2200 & 64 \\
\hline\hline
\end{tabular}
\label{timeseries}
\end{table} 

The data under consideration were gathered between Julian Date 2452847.15 and 2453816.36 (2003 July 26 to 2006 March 21). While the SMEI data extend beyond this date, subsequent data are still undergoing processing, and will be included in a future paper. 
Due to the orbital geometry of the Coriolis satellite, gamma Doradus is observed by Camera \#2 of SMEI for approximately two thirds of the time. Data are observed concurrently by Camera \#1 for approximately thirty days during this time.
The time series analyzed  here consists of three sets of observations, listed in Table \ref{timeseries}.

Point-to-point scatter of the Camera \#1 dataset was estimated at 4.94 parts-per-thousand (ppt), and of the Camera \#2 dataset at 9.27 ppt. When considering the datasets as a whole the Camera \#2 fill was 46.5 per cent, and the Camera \#1 fill was 6.2 per cent.

\section{Analysis}

The data show a large annual periodicity, or `footprint', which has been tentatively associated with changes in the temperature of the camera. This will contribute broad-band noise to the spectrum due to redistribution by irregular gaps in the window of the low-frequency power associated with the footprint. In order to diminish the confusion noise associated with this power a smoothed profile was subtracted from the data. This profile was obtained by a boxcar smoothing of the data, taking a median value within each 10-day box.

Data were analyzed for periodicities with the PERIOD04 package (Lenz and Breger 2005). The periodogram of the whole Camera \#2 dataset is shown in Figure \ref{whole}. Features at the three known frequencies are clearly visible. Also visible are features at 1 day and harmonics of a day, which are a feature of the SMEI data, being associated with  stray-light, and daily periodicities in the level of cosmic rays as the satellite passes though regions variously more or less protected by the Van Allen belts. When considering the gamma Doradus dataset, power at frequencies 1.0, 2.0, 3.0, 4.0 and 5.0 d$^{-1}$, along with additional peaks at 1 d$^{-1}$ $\pm$ (1 year)$^{-1}$, was removed by PERIOD04 using sine-wave fitting in the time domain, a procedure equivalent to the one-dimensional `CLEAN' algorithm (Roberts, 1987).

Canonically a signal-to-noise ratio of 4 in amplitude is taken as indicating that the amplitude seen in a single bin is significant (e.g. Breger et al., 1993). As the Camera \#2 dataset is longer in duration that other available datasets, we checked to ensure that this threshold would be an appropriate measure of significance, testing to see whether a spike at a given signal-to-noise ratio may be reasonably explained as the product of a broadband noise background. This will in general depend upon both the length of dataset, and whether any oversampling has been used, further details of which will be discussed in a forthcoming technical note. Simulations of the Camera \#2 dataset showed that a spike at a S/N of 4.0 has a roughly 5\% chance of occurrence. As the spike of lowest significance we choose to report in this dataset is of S/N 6.2, the chance of occurrence as a produce of noise is very low.

\section{Results}

\begin{figure}
\sidecaption
\includegraphics[width=0.5\textwidth]{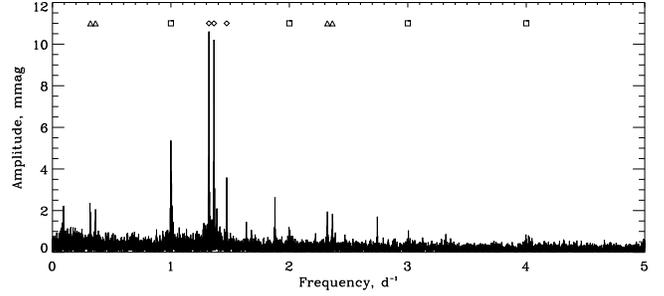}
\caption{Periodogram of the entire Camera \#2 time series. Previously identified frequencies have been marked with diamond symbols, aliases of these with triangles, and daily harmonics with squares.}
\label{whole}
\end{figure} 

\begin{figure}
\centering
\includegraphics[width=0.5\textwidth]{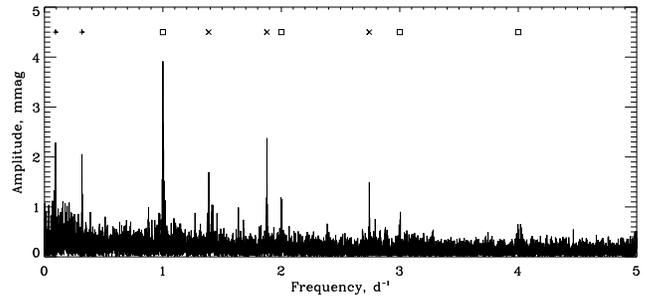}
\caption{Having removed the previously identified frequencies from the Camera \#2 time series, the additional frequencies (marked with 'x' symbols) can be distinguished from the prominent structure of overtones of 1 day, marked with squares. The low frequency spikes, suspected as being non-stellar in origin, have been marked with plus symbols.}
\label{lessknown}
\end{figure}

\begin{figure}
\sidecaption
\includegraphics[width=0.5\textwidth]{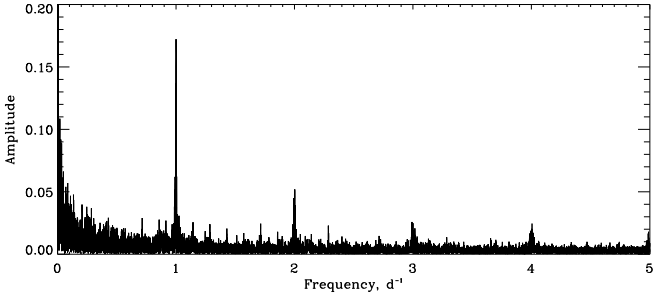}
\caption{The spectral window of the entire Camera \#2 time series; here the zero frequency component has been normalised so as to have a power of 1.0.}
\label{spectralwin}
\end{figure}

The periodogram of the whole Camera \#2 data is seen in Figure \ref{whole}. The three previously identified frequencies (labeled as f1, f2 and f3 in Table \ref{frequencies}), marked with diamond symbols, are clearly visible. Also present are strong aliases of f1 and f2 at $\pm$ 1 d$^{-1}$ (marked by triangles), and the stray light contributions at 1 d$^{-1}$ and harmonics (marked by squares). After pre-whitening in the time-domain to remove the three previously identified frequencies, a further three prominent spikes (f4, f5, f6), marked by `x' symbols in the plots, were observed in the amplitude spectrum (Figure \ref{lessknown}). These spikes do not seem to be associated with the daily harmonics, nor with features of the spectral window (Figure \ref{spectralwin}).

In order to demonstrate thoroughly the robustness of the identification of these additional frequencies, two approaches were taken; first each of the periods of observation in Camera \#2 data highlighted in Table \ref{timeseries} was considered in isolation, and second the Camera \#1 data was analyzed in its entirety.

The significance of features was assessed relative to the noise remaining in the power spectrum after having removed the six noted frequencies, and spikes associated with daily periodicities at 1, 1+(1/year), 1-(1/year), 2, 3, 4,  and 5 d$^{-1}$, and the three additional unidentified frequencies discussed below. A broad range centred on the frequency in question was used to determine the background noise level. In every case the signal-to-noise ratio rises above 4.0, with the exception of f5 in the first observation period of Camera \#2, where the signal to noise is still greater than 3.95 (rounded to 4.0 in Table \ref{frequencies}).

\begin{table*}
\centering
\caption{Parameters related to the observed modes. Errors are based upon the standard deviation of 100 Monte Carlo simulations of the full-length Camera \#2 data. 1, 2 and 3 refer to the periods of observation shown in Table \ref{timeseries}. Notice should be taken that the Balona et al. (1994) amplitudes are in $V$ magnitudes, whereas the SMEI response corresponds best to $R$ magnitudes.}

\begin{tabular}{l|cc|cc|cccc|c}
\hline\hline
	&	\multicolumn{2}{|c|}{Balona et al. (1994)}	&	\multicolumn{2}{|c|}{This Paper}
	&	\multicolumn{5}{|c}{Signal-to-Noise}	\\
\hline
	&	\multicolumn{2}{|c|}{ }	
	&	\multicolumn{2}{|c|}{ }	
	&	\multicolumn{4}{|c|}{Cam \#2}	
	&	Cam \#1	\\
	&	Freq. (d$^{-1}$)	&	Amp. (mmag)	&	Freq. (d$^{-1}$)	&	Amp. (mmag)	&	Whole	&	1	&	2	&	3	&	\\
\hline
f1	&	1.32098(2)	&	11.16(50)
	&	1.32093(2)	&	10.7(2)	
	&	34.3	&	24.6	&	27.1	&	25.1
	&	21.6	\\
f2	&	1.36354(2)	&	12.93(51)
	&	1.36351(2)	&	10.3(3)
	&	33.2	&	22.5	&	26.9	&	24.9
	&	20.4	\\
f3	&	1.475(1)	&	-
	&	1.471(1)	&	3.8(8)
	&	12.5	&	8.0		&	10.6	&	9.6
	&	8.7		\\
f4	&	-	&	-	
	&	1.87(3)	&	2.6(1.1)
	&	8.6		&	5.5		&	7.3		&	7.2
	&	5.8		\\
f5	&	-	&	-
	&	1.39(5)		&	2.0(7)
	&	6.4		&	4.0		&	4.5		&	5.8		&	5.6		\\
f6	&	-	&	-
	&	2.74268(8)		&	1.8(3)
	&	6.8		&	4.6		&	5.6		&	6.1		&	4.1		\\
\hline\hline
\end{tabular}

\label{frequencies}

\end{table*}

Considering the identification of the additional frequencies we speculate that the features at 1.38 and 1.878 d$^{-1}$ represent additional modes of oscillation. The feature at 2.743 d$^{-1}$ occurs notably outside of the frequency range established by the other modes and is at a frequency consistent within the uncertainties with an identification as the linear combination of frequencies f2 + f5.

There are a number of additional features in the periodogram. A notable spike is located at 0.316 d$^{-1}$, at a high amplitude (2.37 mmag) and signal-to-noise ratio (9.5 in the Camera \# 2 dataset), however the signal does not appear at such large significance in the Camera \#1 data. As periodicities greater than 10 days have been suppressed in our analysis, it may be that this feature is part of the rising background noise. Another low-frequency spike occurs at 0.0932 d$^{-1}$, which is almost coincident in frequency with the smoothing box size, and may be an artefact of the smoothing process. A further feature is observed at 1.0829 d$^{-1}$, however as this is equal to 1 d$^{-1}$ + 3$\times$(1 year)$^{-1}$, and the frequency offset, 3$\times$(1 year)$^{-1}$ [= (120 d)$^{-1}$] is comparable in length to the gaps between successive sets of observations, we do not feel confident in ascribing a stellar origin to this spike.

Having cleaned the six frequencies we associate with stellar oscillation, the features associated with daily harmonics, and the further features noted above, no spikes were observed at a S/N of above 4.0 in all tests. However, when considering the full-length Camera \#2 dataset, a number of spikes in the region of 1 to 2 d$^{-1}$ appear at a S/N of greater than 4.0, notably at 1.637 d$^{-1}$, with a S/N of 4.7, leading us to speculate that further modes may be present in gamma Doradus, which may require longer and higher-precision datasets to detect unambiguously. 

\begin{figure}
\includegraphics[width=0.5\textwidth]{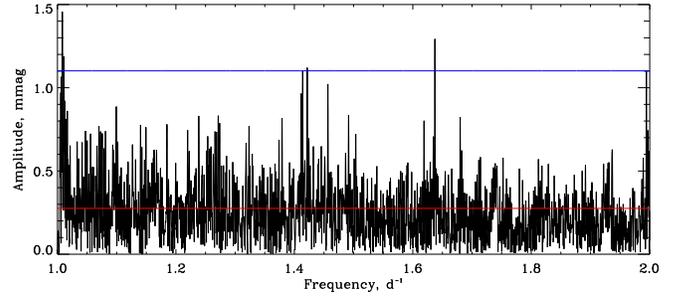}
\caption{A plot of the region between 1 and 2 d$^{-1}$ after the removal of the frequencies, f1 through f6, and daily harmonics, shows the presence of a few additional spikes at around the required S/N threshold (blue), suggesting further modes may be present in the star. }
\label{others}
\end{figure}

\section{Conclusion}

Three additional frequencies of oscillation have been identified in the star gamma Doradus. It is confirmed statistically that they are unlikely to be a product of the noise background. Two of these additional frequencies appear to be oscillations, with a third as a possible combination frequency of a known frequency with a new frequency. The identification is robust, with the frequencies consistently seen across three periods of observation with one camera, and in independent concurrent observations with a second camera.

%Text of your paper...
%... model of the oblique pulsator (Kurtz 1982, Dziembowski \& Goode 1985, Hubrig et al. 2000b, Kurtz et al. 1990, 1996) .....
%\section*{Next Section}
% ... has been done by Dziembowski & Goode (1996), Bigot et al. (2000) ....

%Here comes a figure...
%\figureDSSN{filename}{caption}{label}{!ht}{clip,angle=X,width=XXmm}

%More text...

\begin{acknowledgements}

The authors acknowledge the support of STFC. SMEI was designed and constructed by USAFRL, UCSD, Boston College, Boston University, and the University of Birmingham. NJT would like to thank G. Handler for useful discussions. 
\end{acknowledgements}

\end{document}